\documentclass[doublespacing]{elsart}

\usepackage{amssymb}
\usepackage{epsfig}
\usepackage{tabularx}
\usepackage[english]{babel}
\usepackage{graphicx}
\usepackage[numbers]{natbib}

\begin{document}
\begin{frontmatter}

\title{A highly efficient low-dissipation hybrid weighted essentially non-oscillatory scheme}


\author[label1]{X. Y. Hu}, \author[label2]{B. Wang} \and \author[label1]{N. A. Adams}

\address[label1]{Lehrstuhl f\"{u}r Aerodynamik, Technische Universit\"{a}t
M\"{u}nchen
\\ 85748 Garching, Germany}
\address[label2]{School of Aerospace, Tsinghua University \\
100084 Beijing, China}
\begin{abstract}
In this paper, we propose a simple hybrid WENO scheme
to increase computational efficiency and decrease numerical dissipation.
Based on the characteristic-decomposition approach, 
the scheme switches between the numerical fluxes with characteristic-wise WENO reconstruction
and with the component-wise corresponding optimal linear reconstruction according to a new discontinuity detector.
This non-dimensional discontinuity detector 
measures the resolution limit of the linear scheme
and does not degenerate at critical points. 
A number of numerical examples on inviscid and viscid flow problems 
computed with 5th-order WENO scheme suggest that,
while achieving small numerical dissipation and good robustness, 
the hybrid scheme generally has the total amount calls for the WENO scheme less than 2\%.
\end{abstract}
\end{frontmatter}
\section{Introduction} \label{sec:intro}
High-order weighted essentially non-oscillatory (WENO) schemes \cite{jiang1996cient}
are non-linear schemes which are generally suitable for the simulation of
shock-turbulence interaction due to their high-resolution properties.
However, there are still two important issues
hindering their extensive application.
One is that the cost of computing nonlinear weights 
and implementing the local characteristic decomposition is very high.
The other is that WENO schemes are more dissipative than many low-dissipative linear schemes, particularly
in regions without strong shock wave but with large density variation or shear rates.

One promising approach to overcome these drawbacks is 
using hybrid methods which switch or blend between the numerically dissipative but stable 
non-linear WENO scheme and a less dissipative but more accurate linear scheme 
according to a shock sensor or discontinuity detector
\cite{AdamsShariff96, johnsen2010assessment, pirozzoli2011, li2010hybrid}.
Generally, there are two types of hybrid methods based on the trouble-cell hybridization \cite{AdamsShariff96} 
and the characteristic-wise hybridization \cite{ren2003characteristic}.
In the trouble-cell hybridization, the trouble or discontinuity-crossing cells 
are first detected and tagged by measuring the flow properties such as density, velocity, pressure and entropy. 
Then the numerical fluxes at the cell-faces of trouble cells and their neighboring cells are computed by the WENO scheme.
One issue of this hybridization 
is the lack of smooth transition between the linear and WENO schemes.
This may be due to the suddenly introduced excessive numerical dissipation in the neighbors of trouble cells,
or the very different numerical properties between 
the WENO scheme and the chosen linear scheme \cite{larsson2008stability}. 
Another issue is that the discontinuity detector based on flow properties 
may not be able to detect all discontinuities or multiple discontinuities closely located 
and, hence, lead to numerical instabilities \cite{kawai2008localized}. 
Furthermore, the performance, 
i.e. the overall numerical dissipation, 
numerical stability and computational efficiency,
of the trouble-cell hybridization is strongly dependent on problem, 
grid resolution and discontinuity-detector parameter \cite{pirozzoli2011, li2010hybrid}.
Compared to the trouble-cell hybridization,
since the characteristic-wise hybridization 
switches between linear and WENO schemes  
at each cell face without taking account its neighboring cells and on each characteristic variable,
it provides a much more smoother transition 
and is able to detect all possible discontinuities. 
Also, the discontinuity detector used usually measures 
the less problem-dependent non-dimensional quantities
corresponding to the ratio between the combinations of the approximated low-order derivatives 
\cite{ren2003characteristic, kim2005high},
or the ratio between the approximated high- and low-order derivatives 
\cite{zhou2012family}.
One issue of the characteristic-wise hybridization is that, 
when the solution is near critical points with zero low-order spatial derivatives, 
even the solution is still smooth, the discontinuity detector degenerates 
and switches on the WENO scheme.
To overcome such difficulty, very often multiple dimensional parameters are still tuned 
to obtain more accurate or more stable solution for different problems.
Another issue is that, up to now, 
the computational efficiency of the characteristic-wise hybridization  
usually is much less than that of the trouble-cell hybridization due to 
the large computational effort for the characteristic decomposition
\cite{ren2003characteristic, kim2005high,zhou2012family}.

In this paper, we propose a simple, 
highly efficient hybrid WENO scheme with low dissipation based on
the characteristic-wise hybridization. 
The scheme switches the numerical fluxes of each characteristic variables between those of
the WENO scheme and its corresponding optimal linear scheme without computing the WENO weights. 
Different from previous characteristic-wise hybridization 
\cite{ren2003characteristic, kim2005high, zhou2012family},
the proposed linear scheme is specially designed for much higher efficiency 
by applying component-wise reconstruction
and omitting most of the characteristic-decomposition operations.
The scheme employs a new non-dimensional discontinuity detector, 
which measures the resolvability of the linear scheme.
Compared with previous discontinuity detectors 
\cite{ren2003characteristic, kim2005high, zhou2012family},
since the new detector does not degenerate at critical points 
and it decreases the numerical dissipation and 
improves considerably the accuracy of the solution in the smooth region. 
Furthermore, we show by extensive numerical examples that, by choosing a broadly effective switch threshold,
the hybrid scheme has achieved the total amount calls for the WENO scheme generally less than 2\% without compensating numerical robustness.
\section{Method}\label{sec:method}
We assume that
the fluid is inviscid and compressible, described by the one-dimensional Euler equations as
\begin{equation}\label{governing-equation}
\frac{\partial \mathbf{U}}{\partial t} + \frac{\partial \mathbf{F}(\mathbf{U})}{\partial x} = 0,
\end{equation}
where $\mathbf{U} = (\rho, m, E)^{T}$, and $\mathbf{F}(\mathbf{U}) = [m, \rho u^2 + p, (E + p)u]^{T}$. This set of equations describes the conservation laws for mass density
$\rho$, momentum density $m \equiv \rho u$ and
total energy density $E=\rho e + \rho u^2/2$, where $e$
is the specific internal energy. To close this set of equations, the ideal-gas equation of state $p = (\gamma -1)\rho e$ with a
constant $\gamma$ is used. 
\subsection{Characteristic-wise WENO scheme}
For completeness, we briefly recall the classical 5th-order WENO scheme \cite{jiang1996cient}
for solving Eq. (\ref{governing-equation}). 
The discretization is within the spatial domain such that $x_i = 
i\Delta x$, $i=0, ..., N$, where $\Delta x$ is the spatial step,
the semi-discretized form by the method of lines yields a system of ordinary differential equations
\begin{equation}\label{conservative-scheme}
\frac{d \mathbf{U}_i}{dt} = \frac{1}{\Delta x} \left(\hat{\mathbf{F}}_{i-1/2} - \hat{\mathbf{F}}_{i+1/2}\right),
\end{equation}
where $\hat{\mathbf{F}}_{i\pm1/2}$ are the numerical flux at $x_{i\pm1/2}$, respectively.
Once the right-hand side of this expression has been evaluated, a TVD
Runge-Kutta method is employed to advance the solution in time.
In the typical characteristic-wise finite-difference WENO scheme,
the $\hat{\mathbf{F}}_{i+1/2}$ are usually reconstructed within the local characteristic fields. 
Let us take the matrix $\mathbf{A}_{i+1/2}$ to be 
the Roe-average Jacobian $\partial \mathbf{F} /\partial \mathbf{U} $ at $x_{i+1/2}$.
We denote by $\lambda_s$, $\mathbf{r}_s$ (column vector) and $\mathbf{l}_s$ (row vector) 
the $s$th eigenvalue, and the right and left eigenvectors of $\mathbf{A}_{i+1/2}$, respectively.
The physical fluxes and conservative variables on the respective reconstruction stencil 
are mapped to the characteristic field by the characteristic-projection step 
\begin{equation}\label{projection}
v_{j,s} = \mathbf{l}_s \cdot \mathbf{U}_j, \quad g_{j,s} = \mathbf{l}_s \cdot \mathbf{F}_j,
\end{equation}
where $i+3>j>i-2$ and $s= 1, 2, 3$.
For each component of the characteristic variables, 
the corresponding split numerical fluxes are constructed by
\begin{equation}\label{split-flux}
f^{+}_{j,s} = \frac{1}{2}\left(g_{j,s} + \alpha_s v_{j,s}\right), \quad 
f^{-}_{j,s} = \frac{1}{2}\left(g_{j,s} - \alpha_s v_{j,s}\right),
\end{equation}
where $\alpha_s  = |\lambda_s|$ for a Roe flux (RF).
Alternatively one can use $\alpha_s =\max|\lambda_{l,s}|$, 
where $l$ represents the entire computational domain for a Lax-Friedrichs flux (LF)
or the neighborhood of $i$ for a local Lax-Friedrichs flux (LLF).
The WENO reconstruction gives 
\begin{equation}\label{WENO}
f^{+}_{i+1/2,s} = \sum^{2}_{k=0}\omega^{+}_{k,s}
f^{+}_{k, i+1/2},    \quad f^{-}_{i+1/2,s} = \sum^{2}_{k=0}\omega^{-}_{k,s}
f^{-}_{k, i+1/2,s},
\end{equation}
where $f^{\pm}_{k, i\pm1/2,s}$ are 3rd-order reconstructions from upwind 3-point stencils,
and $\omega^{\pm}_{k,s}$ are WENO weights defined by
\begin{equation}
\omega^{\pm}_{k,s} = \frac{\alpha_{k}}{\sum^{2}_{k=0}\alpha_{k,s}},
\quad \alpha_{k,s} = \frac{d_k}{\left(\beta_{k,s} + \epsilon\right)^q},
\label{weight}
\end{equation}
where $d_0 = \frac{3}{10}$, $d_1 = \frac{3}{5}$ and $d_2 = \frac{1}{10}$ are optimal weights. 
These optimal weights generate the 5th-order upwind scheme, 
by which the numerical flux is reconstructed from a 5-point stencil.
$\epsilon>0$ prevents division by zero, 
$q = 1$ or $2$ is chosen to adjust the distinct weights,
and $\beta_{k,s}$ are the smoothness indicators. 
We refer to Jiang and Shu \cite{jiang1996cient} for the details of WENO reconstruction. 
The numerical flux in each characteristic field is then computed by 
\begin{equation}\label{un-split-flux}
\hat{f}_{i+1/2,s} = f^{+}_{i+1/2,s}  + f^{-}_{i+1/2,s}.
\end{equation}
At last, this numerical flux is projected back to the physical space by
\begin{equation}\label{component-flux}
\hat{\mathbf{F}}_{i+1/2} =\sum^{3}_{s=1}  \hat{f}_{i+1/2,s}\mathbf{r}_s.
\end{equation}
It can be found that the major part of floating-point operations in 
the characteristic-wise WENO scheme are due to 
the characteristic-projection step (matrix-vector product) of Eq. (\ref{projection}),
and the computation of the WENO weights in Eq. (\ref{weight}).
\subsection{Hybrid WENO scheme}
In order to decrease the number of floating-point operations, 
we propose to hybridize the WENO scheme with its optimal linear scheme
by the characteristic-decomposition approach.
A straightforward implementation is to switch between Eq. (\ref{un-split-flux}) 
and the corresponding flux of the optimal linear scheme based on the characteristic-variable projection. 
Here, however, we consider a component-wise reconstruction for the optimal linear scheme
to further save the computational effort for the characteristic-projection step.

\subsubsection{Optimal linear scheme with component-wise reconstruction}
In this formulation, the approximated physical fluxes and the differences of approximated conservative variables 
at $x_{i+1/2}$ are computed component-by-component as
\begin{equation}\label{cell-wall-component-flux}
\mathbf{F}_{i+1/2} = \frac{1}{2}\left(\mathbf{F}^{+}_{i+1/2} + \mathbf{F}^{-}_{i+1/2}\right),
\quad \Delta \mathbf{U}_{i+1/2} = \frac{1}{2}\left(\mathbf{U}^{+}_{i+1/2} - \mathbf{U}^{-}_{i+1/2}\right).
\end{equation}
With the optimal linear scheme of the 5th-order WENO scheme, one has
\begin{eqnarray}
\mathbf{F}_{i+1/2} & = & \frac{1}{60}\left(\mathbf{F}_{i-2} - 8\mathbf{F}_{i-1} + 37\mathbf{F}_{i} + 37\mathbf{F}_{i+1} - 8\mathbf{F}_{i+2} + \mathbf{F}_{i+3}\right), \label{cell-wall-interpolation-1}\\
\Delta \mathbf{U}_{i+1/2} & = & \frac{1}{60}\left(\mathbf{U}_{i-2} - 5\mathbf{U}_{i-1} + 10\mathbf{U}_{i} - 10\mathbf{U}_{i+1} + 5\mathbf{U}_{i+2} - \mathbf{U}_{i+3}\right) \label{cell-wall-interpolation-2}.
\end{eqnarray}
Note that Taylor-series expansion of Eq. (\ref{cell-wall-interpolation-2}) 
suggests 
\begin{equation}\label{accuracy}
\Delta \mathbf{U}_{i+1/2} = -\frac{\Delta x^{5}}{60}\left.\frac{\partial^{5} \mathbf{U}(x)}{\partial x^{5}}\right|_{i+1/2} + O(\Delta x^{7}).
\end{equation}
Then $\mathbf{F}_{i+1/2}$ and $\Delta \mathbf{U}_{i+1/2}$ are projected onto the characteristic field by 
\begin{equation}\label{projection-linear}
\Delta v_{i+1/2,s} = \mathbf{l}_s \cdot \Delta \mathbf{U}_{i+1/2}, \quad g_{i+1/2,s} = \mathbf{l}_s \cdot \mathbf{F}_{i+1/2}.
\end{equation}
For each component of the characteristic variables, the corresponding numerical flux
is constructed by
\begin{equation}\label{split-flux-linear}
\hat{f}_{i+1/2,s} = g_{i+1/2,s} + \alpha'_s \Delta v_{i+1/2,s},
\end{equation}
where $\alpha'_s = |\lambda_s|$. 
Note that such choice of RF fluxes leads to less numerical dissipation compared to that of LLF or LF fluxes. 
At last, the numerical flux obtained in each characteristic field 
is projected back to the component space by Eq. (\ref{component-flux}). 
Note that, compared to the characteristic-wise WENO scheme,
the linear scheme omits the computation of the WENO weights in Eq. (\ref{weight}),
and decreases by $5/6$ the characteristic-projection operations of Eq. (\ref{projection}).
Also note that such properties of the present single-component characteristic-projection
allows the hybridization applied within the characteristic field, as shown in the next section, 
as in Refs. \cite{ren2003characteristic, kim2005high, zhou2012family} with much less computational effort.
\subsubsection{Hybridization and discontinuity detector}
Since the linear scheme is numerically unstable for solutions with discontinuities,
it should be switched on only in smooth regions of the solution 
according to a discontinuity detector. 
A less problem-independent discontinuity detector than those in the 
trouble-cell hybridization 
\cite{AdamsShariff96, pirozzoli2002conservative, li2010hybrid}
are usually based on non-dimensional measurement on the characteristic variables.
A usual way to achieve this is using the ratio between 
low order undivided differences \cite{ren2003characteristic, kim2005high},
or the ratio between high and low order undivided differences \cite{li2010hybrid}.
Since undivided difference corresponds to approximation of a spatial derivative,
this type of discontinuity detectors degenerate near critical points 
with zero low order derivatives as the denominators approach zero 
and switches on the WENO scheme. 
To decrease the over-dissipation associated with the degeneration, 
dimensional parameters must be introduced and tuned for different problems. 

In this paper, we achieve a general effective measure in a different way.
By noticing that the characteristic variables have the dimension of density
we define a non-dimensional discontinuity detector by
\begin{equation}\label{indicator}
\sigma_{s} = \left(\frac{\Delta v_{i+1/2,s}}{\tilde{\rho}}\right)^2,
\end{equation}
where $\tilde{\rho}$ is  the Roe-average density of $\mathbf{A}_{i+1/2}$.
Note that there may be different forms of left and right eigenvectors 
which lead to different dimension of the characteristic variables.
In this case, the expected dimension can be achieved by multiplying
or diving a factor to corresponding eigenvectors.
From Eq. (\ref{accuracy}) it can be found that 
$\sigma_{s}$ also is related to the 5th-order derivatives of the characteristic variables 
due to the linearized characteristic projection.
Since the optimal 5th-order linear scheme reconstructs with 4th-degree polynomials
and is not able to resolve 5th or higher order derivatives, 
$\sigma_{s}$ is actually a measure on the resolvability of the linear scheme.
If there is no vacuum in the flow (which is almost always the case), 
the denominator of $\sigma_{s}$ does not degenerate due to the finite value of $\tilde{\rho}$.

With a given threshold expression 
\begin{equation}\label{threshold}
\varepsilon = C \left(\frac{\Delta x}{L}\right)^{p} \ll 1,
\end {equation}
where $C>0$ is a constant, $L$ is the characteristic length scale of the problem and 
$p>0$ is an integer,   
the hybrid scheme can be constructed as follows:
for each component of the characteristic variables
the numerical flux is obtained by linear flux of Eq. (\ref{split-flux-linear}) if $\sigma_{s} < \varepsilon$;
otherwise it is obtained by WENO flux of Eq. (\ref{un-split-flux}).
Note that neither Eq. (\ref{indicator}) or Eq. (\ref{threshold}) introduces any dimensional parameter.
As will be shown in the next section also, 
$(\Delta x/L)^{p}$ is introduced to achieve convergence for capturing discontinuities. 
Assume Eq. (\ref{indicator}) measures on the cell-face at $i-1/2$ next to a discontinuity at $i+1/2$,
the hybrid scheme tolerates, i.e. by applying the optimal linear scheme, 
a small overshoot or undershoot scaling with $\varepsilon$. 
Due to the self-similar property of the discontinuity, $\sigma_{s}$ is unchanged with increasing resolution.
In this case, convergence requires that $\varepsilon$ decreases, where $p$ determines the rate, 
with increasing resolution.

Alo note that the above approach can be directly applied to multiple dimensions by a dimension-by-dimension approach, 
and can be very easily extended to higher-order or modified WENO schemes.
The only modification for other WENO schemes is that 
Eq. (\ref{cell-wall-component-flux}) should be computed with the corresponding optimal linear schemes. 
\section{Numerical examples}
The following numerical examples are provided to illustrate the
potential of the proposed hybrid WENO scheme.  
The governing equations are the one- and two-dimensional compressible Euler or Navier-Stokes equations.
While the original 5th-order WENO is denoted as WENO-5,
the present hybrid scheme is denoted as H-WENO-5.
The 3rd-order TVD Runge-Kutta scheme is used for time integration \cite{shu1989efficient}. 
In order to achieve an objective measurement on the computational cost, 
the local efficiency on each cell-face through the entire computation time is defined by
$\eta = M_{Linear}/(M_{Linear} + M_{WENO})$, 
where $M_{WENO}$ is the number of operations with WENO flux and 
$M_{Linear}$ is the number of operations with linear flux.
The overall computational efficiency is obtained by averaging $\eta$
over the entire computational domain.
In order to show the general effective discontinuity detector of Eq. (\ref{indicator}), 
We set the parameters $C = 100$ and $p=3$  in Eq. (\ref{threshold}) for all numerical examples 
(see also a study of the sensitivity of $C$ in Sec. \ref{sec:blast-wave}). 
If not mentioned otherwise, 
all the computations are carried out with a CFL number of 0.6.
\subsection{Propagation of broadband sound waves}
This problem, taken from Sun et al. \cite{sun2011class}, corresponds to the propagation of a sound wave packet which contains acoustic turbulent structure with various length scales. The initial condition is
\begin{eqnarray} \label{acoustic}
p(x,0) & = & p_0\left\{ 1 + \epsilon \sum^{N/2}_{k=1}\left[E_p(k)\right]^{1/2}\sin\left[2\pi k(x + \phi_k)\right]\right\},
\nonumber \\
\rho(x,0) & = & \rho_0\left[\frac{p(x,0)}{p_0}\right]^{1/\gamma}, \nonumber\\
u(x,0) & =  & u_0 + \frac{2}{\gamma-1}\left[\frac{c(x,0)}{c_0}\right], \nonumber
\end{eqnarray}
where $\phi_k$ is a random number between 0 and 1 with uniform distribution,
$\epsilon= 10^{-3}$, $\gamma = 1.4$, $c$ is the speed of sound and
$$
E_p(k) = \left(\frac{k}{k_0}\right)^4 \exp^{-2(k/k_0)^2}
$$
is the energy spectrum which reaches its maximum at $k = k_0$.
A periodic boundary condition is applied at $x=0$ and $x=1$.
Computations have been carried out on a $128$-point grid 
using a CFL number of 0.2 for one period of time.
\begin{figure}[p]
\begin{center}
\includegraphics[width=1.2\textwidth]{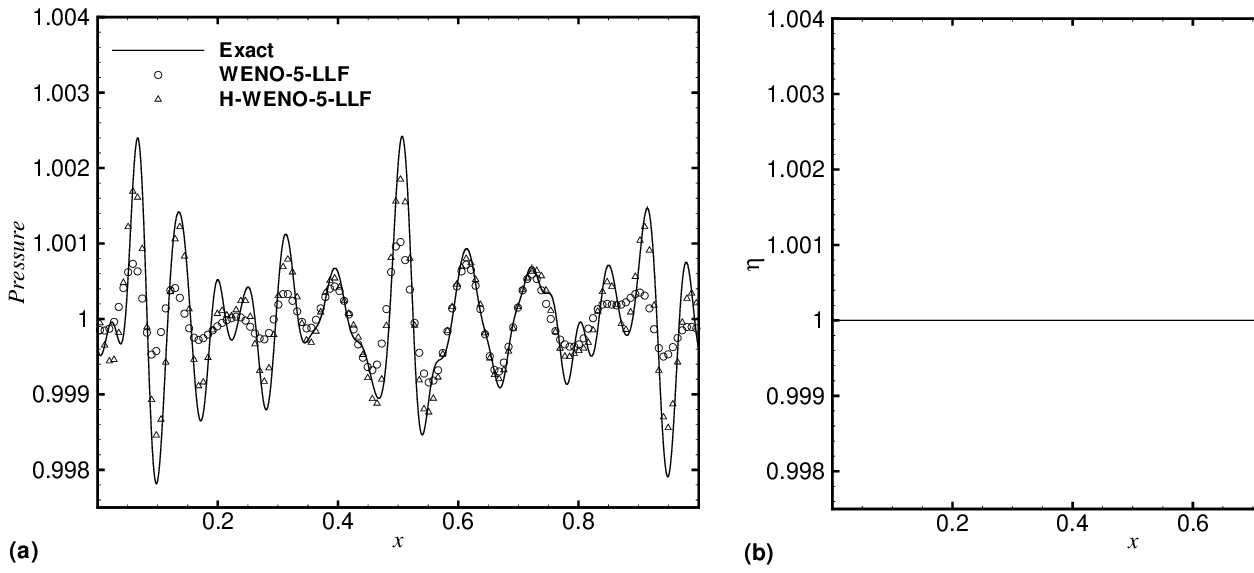}
\caption{Propagation of broadband sound waves computed on a 128 points grid: 
(a) pressure and (b) local computational efficiency distribution.} 
\label{broad-sound}
\end{center}
\end{figure}
The numerical results in Fig. \ref{broad-sound}a show 
that the present methods gives considerably better resolved sound waves than WENO-5, 
especially for the regions near the critical points, 
at which previous discontinuity detectors, such as that used in Sun et al. \cite{sun2011class}, 
can switch on WENO scheme if no dimensional parameter is introduced.
As shown in Fig. \ref{broad-sound}b, 
no WENO flux is switched on during the entire computation,
indicating that the present hybrid scheme recovers the optimal linear scheme, 
therefore is able to achieve the formal 5th-order accuracy even near the critical points, 
where WENO-5 degenerates the order of accuracy and introduces extra numerical dissipation.
\subsection{Shock-tube problems}
Here, we show that the proposed scheme H-WENO-5 passes the shock-tube test
problems: the Sod problem (Sod 1978), the Lax problem (Lax 1954) 
and the 1-2-3 problem (Einfeldt et al. 1991).
For the Sod problem, the
initial condition is
$$
(\rho, u, p)=\cases{(1, 0, 1) & if $0<x<0.5$ \\
\cr (0.125, 0, 0.1) & if $1>x>0.5$ \\},
$$
and the final time is $t=0.2$. For the Lax problem, the initial condition is
$$
(\rho, u, p)=\cases{(0.445, 0.698, 0.3528) & if $0<x<0.5$ \\
\cr (0.5, 0, 0.5710) & if $1>x>0.5$ \\},
$$
and the final time is $t=0.14$. For the 123 problem, the initial condition is
$$
(\rho, u, p)=\cases{(1, -2, 0.4) & if $0<x<0.5$ \\
\cr (1, 2, 0.4) & if $1>x>0.5$ \\},
$$
and the final time is $t=0.1$. 
We examine the numerical solution with difference resolutions to study the performance of the present method with increasing resolution.
\begin{figure}[p]
\begin{center}
\includegraphics[width=1.2\textwidth]{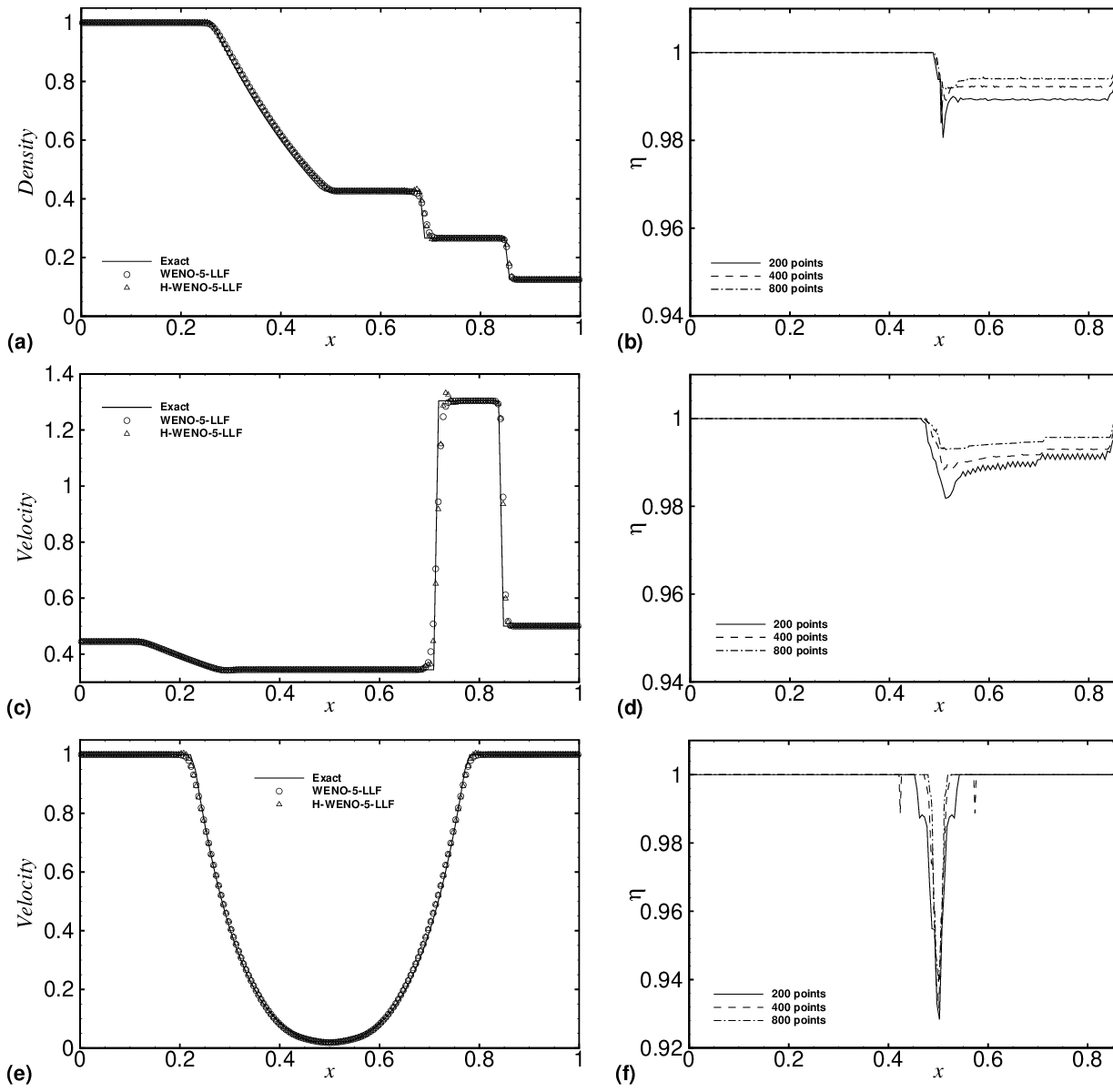}
\caption{Shock-tube problems:
(a, b) Sod problem; (c, d) Lax problem; (e, f) 1-2-3 problem.} 
\label{shock-tube}
\end{center}
\end{figure}
Figures \ref{shock-tube}a, c and e gives the
density distributions computed on a 200-point grid
and local computational efficiency for three grid resolutions.
It can be observed that good agreement 
between the present solution and that of WENO-5. 
As shown in Fig. \ref{shock-tube}b, d and f, 
the computational efficiency increases with resolution.
It can also be observed that the WENO-5 is essentially not switched on in the smooth regions,
even those with 1st or high-order singularities near the fringe of expansion waves. 
For all three problems, 
the overall computational efficiency achieves 99.1\% for the lowest resolution 
and 99.7\% for the highest resolution.
Note that, the computational efficiency achieved here is even much higher 
than those obtained by the trouble-cell hybridization 
using several different discontinuity detectors \cite{li2010hybrid}.
Also note that the hybrid scheme introduces very small overshoots near the discontinuities, 
as shown in Fig. \ref{shock-tube}a and c. 
If the threshold is chosen in the form of Eq. (\ref{threshold}),
as shown by Fig. \ref{sod-convergence},
\begin{figure}[p]
\begin{center}
\includegraphics[width=1.2\textwidth]{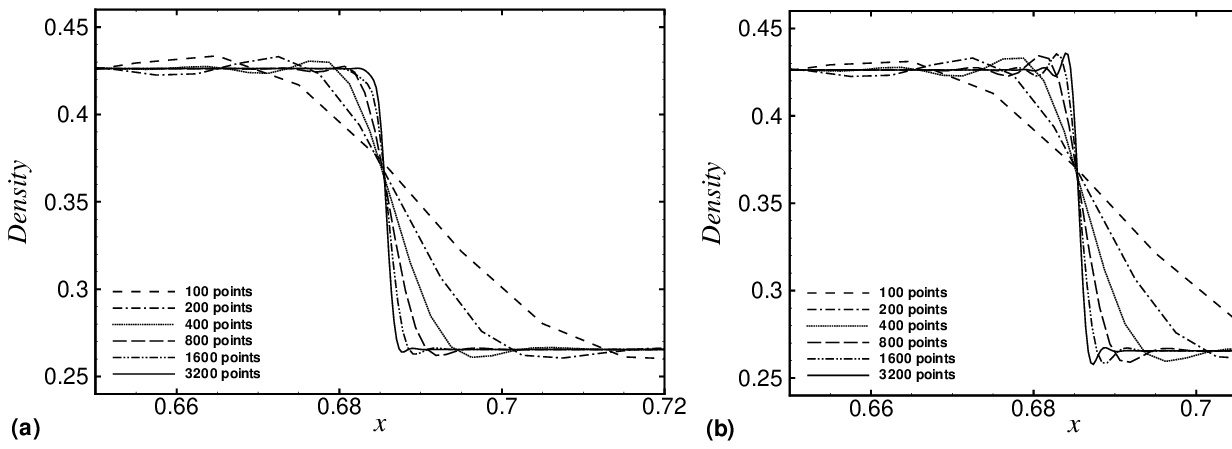}
\caption{Sod problem computed with (a) $\varepsilon = 100 (\Delta x/L)^{3}$ and 
(b) $\varepsilon = 1.25\times 10^{-5}$ for different grid resolutions.} 
\label{sod-convergence}
\end{center}
\end{figure}
these overshoots vanish with increasing resolution and 
the captured discontinuities converge to their exact solution.
However, if a constant threshold is used, 
these overshoots remain at approximately 
the same magnitude therefore no convergence is achieved. 
\subsection{Interacting blast waves}\label{sec:blast-wave}
We consider a two-blast-wave
interaction problem, which is taken from Woodward and Colella \cite{colella1984piecewise} .
The initial condition is
$$
(\rho, u, p)=\cases{(1, 0, 1000) & if $0<x<0.1$ \\ \cr
(1, 0, 0.01) & if $0.1<x<0.9$ \\
\cr (1, 0, 100) & if $1>x>0.9$
\\},
$$
and the final time is $t=0.038$. The reflective boundary condition is applied at both $x=0$ and
$x=1$. We examine the numerical solution on a 400-point grid with the parameters 
$C=$ $10$, $100$ and $1000$.
The reference "exact" solution is a high-resolution solution
on a $3200$-point grid computed by WENO-5.
\begin{figure}[p]
\begin{center}
\includegraphics[width=1.2\textwidth]{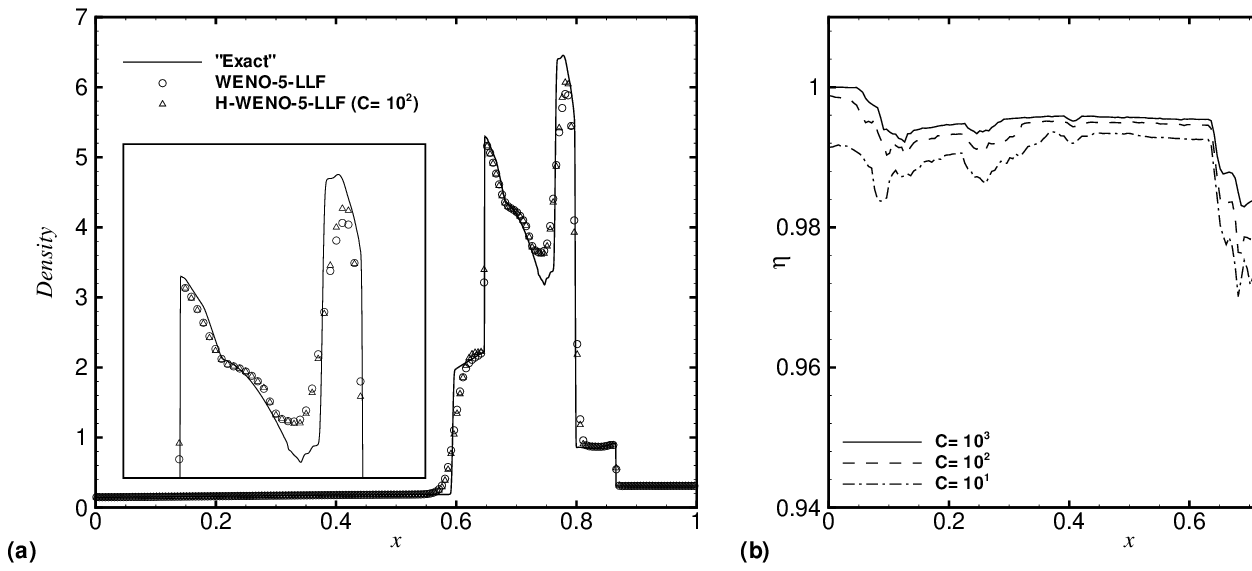}
\caption{Interacting blast waves:
(a) density profile with a closer view; (b) local computational efficiency.} \label{blast}
\end{center}
\end{figure}
Figure \ref{blast} give the profiles of density and local computational efficiency.
Again, good agreement with the reference solutions is observed 
and the computational efficiency is much higher than 
those obtained by the trouble-cell hybridization \cite{li2010hybrid}. 
Due to smaller numerical dissipation H-WENO-5 shows an slightly improved solution compared to WENO-5.
As shown in Fig. \ref{blast}b, 
the computational efficiency is only weakly dependent on the choice of $C$.
Actually, by increasing $C$ 100 times, 
the extra gain of overall efficiency is less than 0.7\% (from 98.3\% to 99.0\%). 
Note that, since the computation (results not shown here) is still stable even with $C=10^{5}$,
the numerical robustness of the present method is also weakly depends on the choice of $C$.
\subsection{Shock-density wave interaction} \label{sec:shock-density}
We consider a shock density-wave interaction problem \cite{shu1989efficient}.
The initial condition is set by a Mach 3 shock interacting with a perturbed density field
$$
(\rho, u, p)=\cases{(3.857, 2.629, 10.333) & if $0\leq x<1$ \\
\cr (1+0.2\sin(5x), 0, 1) & if $10\geq x>1$
\\}
$$
and the final time is $t=1.8$. 
A zero-gradient boundary condition is applied at $x = 0$ and $10$. 
We examine the numerical solution on 200-point and 400-point grids, 
and the reference "exact" solution is a high-resolution solution on 
a $3200$-point grid computed by WENO-5.
\begin{figure}
\begin{center}
\includegraphics[width=1.2\textwidth]{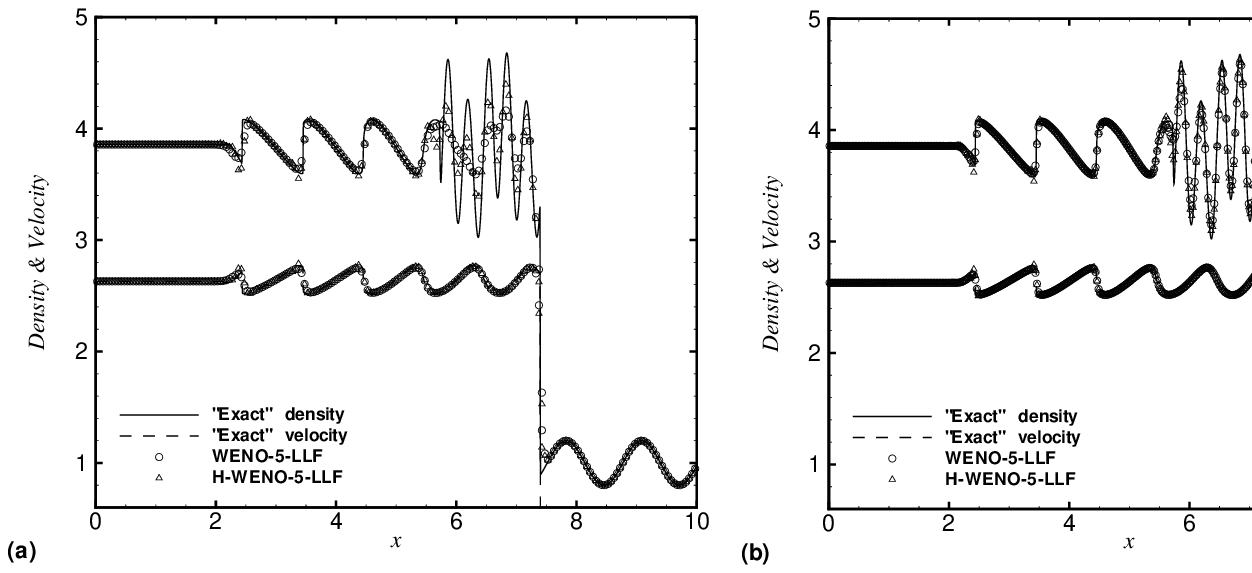}
\end{center}
\caption{Shock-density-wave interaction: 
density and velocity profiles on (a) a 200-point grid and (b) a 400-point grid.}
\label{shock-density}
\end{figure}
Fig. \ref{shock-density}a show the calculated density and
velocity profile.
It can be observed that, when the grid resolution is low, 
H-WENO-5 gives considerably better resolved density waves 
behind the shock than WENO-5, which is too dissipative. 
By considering all cases discussed in previous sections, 
it can be find that the present method gains more significant improvement  
in smoothed region than near the discontinuity.   
\subsection{Double Mach reflection a strong shock}
We consider the problem from Woodward and Colella  \cite{colella1984piecewise} 
on the double Mach reflection of a strong shock.
A Mach 10 shock in air is reflected from the wall with incident angle of $60^0$. 
For this case, the initial condition is
$$
(\rho, u, v, p)=\cases{(1.4, 0, 0, 1) & if $y < 1.732(x - 0.1667)$ \\
\cr (8, 7.145, -4.125, 116.8333) & else \\
},
$$
and the final time is $t = 0.2$. 
The computational domain of this problem is $[0,0]\times[4,1]$.
Initially, the shock extends from the point $x = 0.1667$ at the bottom
to the top of the computational domain.
Along the bottom boundary, at $y = 0$, the region from $x = 0$ to $x = 0.1667$
is always assigned post-shock conditions, whereas a reflecting wall condition
is set from $x = 0.1667$ to $x = 4$. Inflow and outflow boundary conditions
are applied at the left and right ends of the domain, respectively.
The values at the top boundary are set to describe the exact motion of a Mach 10 shock.
We examine the numerical solution with three resolutions on grids of 
$240\times 60$, $480\times 120$ and $960\times 240$ points.

\begin{figure}[p]
\begin{center}
\includegraphics[width=1.2\textwidth]{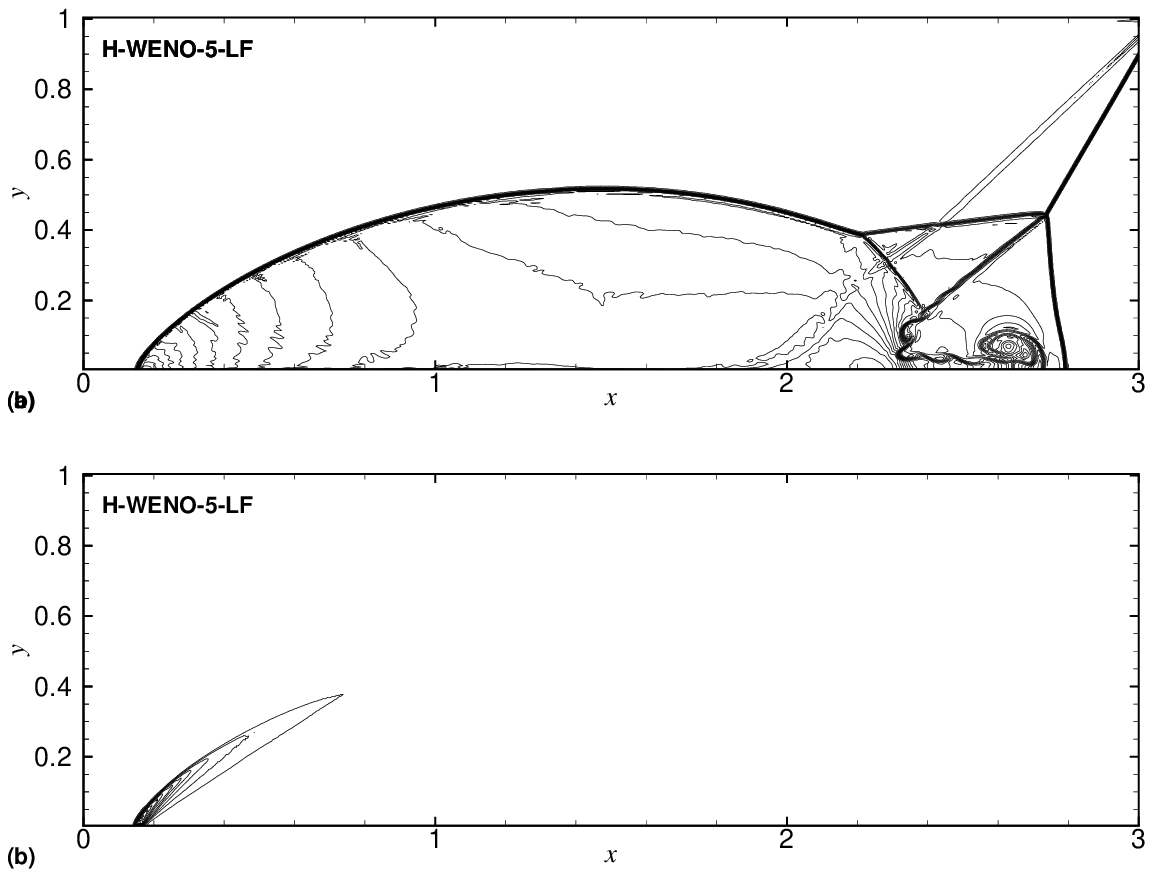}
\caption{Double-Mach reflection of a Mach 10 shock wave at $t = 0.2$ computed on $960\times 240$ grid: (a) 40 density contours from 1.88783 to 20.9144,
(b) 40 contours of local computational efficiency from 0.5 to 1. 
Note that in the most of the computational domain 
the calls for WENO-5 scheme are very few, 
i.e. less than 1.3\% which gives the first contour in (b).} 
\label{double-mach}
\end{center}
\end{figure}

Figure \ref{double-mach} shows  contours of density
and local computational efficiency computed on a $960\times 240$ grid.
It is observed a good agreement with the results of Kim and Kwon \cite{kim2005high} (their Figs. 12 and 13)
computed with fine-tuned parameters at the same resolution.
Compared to the results in Zhou et al. \cite{zhou2012family} 
computed with fine-tuned parameters and higher grid resolution,
much more fine-scale structures are obtained by the present scheme.
Note that, as shown in Fig. \ref{double-mach}b, 
the WENO flux is seldom calculated in the entire computation domain,
except near the end of the main reflection wave. 
The overall computational efficiencies respectively for three resolutions are 
$99.5\%$, $99.6\%$ and $99.7\%$, 
which decreases with increasing grid resolution.
Such computational efficiencies indicate that, 
the computational cost of the WENO scheme is essentially negligible.
\subsection{Viscous shock tube problem}
We consider the two-dimensional viscous flow problem 
in a square shock tube with unit height and
insulated walls \cite{daru2000evaluation}. 
In this problem, the propagation of the incident shock wave 
and contact discontinuity lead to a thin boundary layer. 
After its reflection on the right wall, 
the shock wave interacts with this boundary layer and results a separation region and  
the formation of  a typical "$\lambda$-shape like shock pattern". 
The initial condition is
$$
(\rho, u, p)=\cases{(120, 0, 120/\gamma) & if $0\leq x<\frac{1}{2}$ \\
\cr (1.2, 0, 1.2/\gamma) & if $1\geq x>\frac{1}{2}$
\\}
$$
The fluid is assumed as ideal gas with $\gamma=1.4$ and constant dynamics viscosity $\mu = 0.005$ 
and Prandtl number $Pr= 0.73$ and satisfying the Stokes assumption.
If the reference values are chosen as the initial speed of sound, 
unit density and unit length, the Reynolds number is 200.
By applying the symmetry condition at the upper boundary, only the lower half domain is actually computed.
For other boundaries, the no-slip and adiabatic wall conditions are applied. 
The viscous and heat transfer is calculated by 6th-order accuracy,
Here, we examine the numerical solution with two resolutions on grids of 250$\times$125 and 500$\times250$ points.

\begin{figure}[p]
\begin{center}
\includegraphics[width=1.2\textwidth]{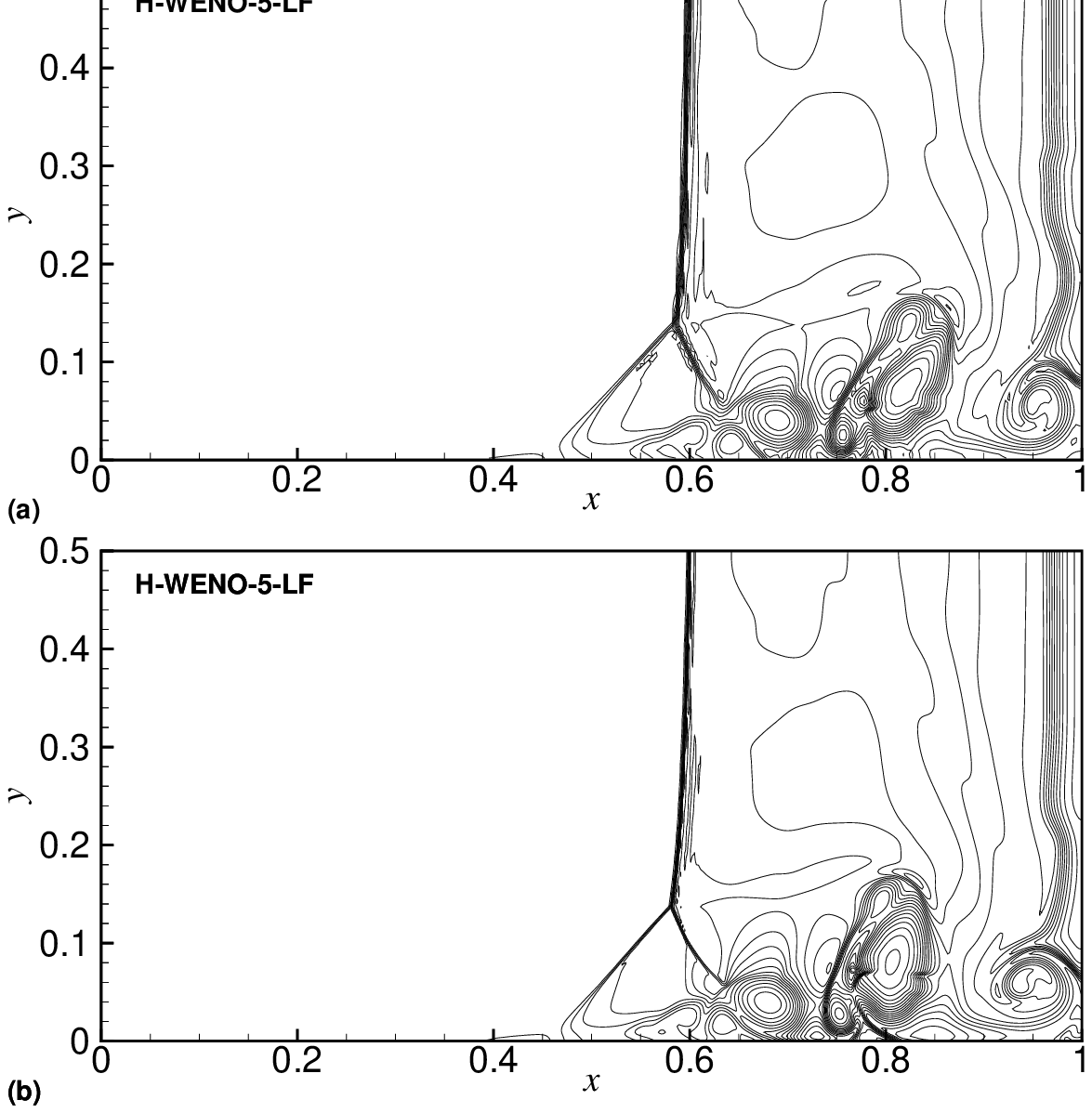}
\caption{Viscous shock tube problem at $t=1$: 20 density contours from 15 to 125 computed on 
(a) a $250\times 125$ grid and (b) a $500\times 250$ grid.} 
\label{vis-shock}
\end{center}
\end{figure}

As shown by the density contours on the coarse grid in Fig. \ref{vis-shock}a, the computed result, 
especially with respect to the shape and the tilting angle of the primary vortex, 
is already considerably better than those obtained by the WENO-5, 
the 6th-order artificial compression method (ACM) with wavelet based filter scheme 
\cite{sjoegreen2003grid} (their Fig. 4) and the 5th-order multi-dimensional limiting process (MLP)
cocombined with M-AUSMPW+ scheme \cite{kim2005accurate} (their Fig. 41) with the same grid resolution.
Grid convergence is indicated for the results computed on the fine grid
because no notable differences, except the near shock region, 
can be identified upon comparison with the grid-converged density contours 
of Sj\"{o}green and Yee \cite{sjoegreen2003grid} (their Fig. 2) obtained with much higher grid resolutions.
Note that there are small disturbances near the shock wave in the low resolution results (see. Fig. \ref{vis-shock}a).
These small disturbances are numerical errors rather than instabilities, 
since they do not propagate to other regions and their magnitudes decreases by increasing resolution (see. Fig. \ref{vis-shock}b).
Also note that the obtained overall computational efficiency for the two resolutions are $99.8\%$, $99.9\%$,
which again imply that the WENO flux is actually hardly utilized throughout the entire computation.
\section{Concluding remarks}
We propose a simple hybrid WENO scheme
to increase computational efficiency and decrease numerical dissipation.
Based on characteristic-wise hybridization, 
the scheme switches the numerical flux of each characteristic variables
between that of the WENO scheme and its optimal linear scheme
according to a discontinuity detector measuring the resolvability of the linear scheme. 
As shown by a number of numerical examples 
the overall computational efficiency, measured by the fraction of linear flux used, 
is always higher than $98\%$, 
shows that the hybridization increases the computational efficiency greatly.
Also, by choosing a general effective threshold for the discontinuity detector,
compared to the original WENO scheme, 
considerable lower numerical dissipation is achieved in smooth region 
without compensating the robustness for capturing discontinuities.
Note that the method can be applied for multiple-species flows with
even higher computational efficiency
due to the serious overhead of local characteristic projection 
and WENO weights calculation for each species.
\bibliographystyle{plain}

\end{document}